\documentstyle[12pt,epsfig]{article}
\def\la{\langle}\def\ra{\rangle}
\def\be{\begin{eqnarray}}\def\bea{\begin{eqnarray}}
\def\ba{\begin{eqnarray}}
\def\ee{\end{eqnarray}}\def\eea{\end{eqnarray}}
\def\ea{\end{eqnarray}}
\def\ben{\begin{eqnarray}}\def\bitem{\begin{itemize}}
\def\een{\end{eqnarray}}\def\eitem{\end{itemize}}

\def\G0p{$G_0^\prime$}
\def\bi{\bibitem}

\def\N1520{$N^\star (1520)$}
\def\calL{{\cal L}}
\def\calO{{\cal O}}

\def\prl{Phys. Rev. Lett.}\def\pr{Phys. Rev.}\def\np{Nucl. Phys.}
\def\pl{Phys. Lett.}
\def\B#1{{}^{#1}\mbox{B}}

\def\roughly#1{\mathrel{\raise.3ex\hbox{$#1$\kern-.75em%
\lower1ex\hbox{$\sim$}}}}\def\lsim{\roughly<}
\def\gsim{\roughly>}

\def\B#1{{}^{#1}\mbox{B}}

\def\itt{\indent\indent}

\def\B0{\mbox{\boldmath $0$}}

\def\bq{\begin{equation}}
\def\eq{\end{equation}}

\def\rhosobar{$\rho_{sobar}$}\def\rhoelm{$\rho_{elm}$}
\def\b3{\mbox{\boldmath $3$}}\def\b6{\mbox{\boldmath $6$}}

\renewcommand{\thefootnote}{\fnsymbol{footnote}}
\begin{document}
\begin{titlepage}
\begin{center}

{\Large \bf  Matching the QCD and Hadron Sectors and Medium
Dependent Meson Masses; Hadronization in Relativistic Heavy Ion
Collisions}
  \vskip 1.2cm
   {{\large G.E. Brown$^{(a)}$ and Mannque Rho$^{(b,c)}$} }
 \vskip 0.5cm

{(a) \it Department of Physics and Astronomy, State University of
New York,\\ Stony Brook, NY 11794, USA}

 {(bc) \it  Service de Physique Th\'eorique, CEA/DSM/SPhT,
Unit\'e de recherche associ\'ee au CNRS, CEA/Saclay,  91191
Gif-sur-Yvette c\'edex, France}

{(c) \it School of Physics, Korea Institute for Advanced Study,
Seoul 130-722, Korea}

\end{center}

\vskip 0.5cm

\centerline{(\today)}
 \vskip 1cm

\centerline{\bf Abstract}
 \vskip 0.5cm
The recent developments on the ``vector manifestation" of chiral
symmetry by Harada and Yamawaki provide a compelling evidence for,
and ``refine," the in-medium scaling of hadronic properties in
dense/hot matter (call it ``BR scaling") proposed by the authors
in 1991. We reinterpret the Harada-Yamawaki result obtained in a
Wilsonian renormalization-group approach to hidden local symmetry
theory matched to QCD at near the chiral scale in terms of the
Nambu-Jona-Lasinio model and predict that the vector meson mass in
medium should scale $m_\rho^\star/m_\rho\sim
(\la\bar{q}q\ra^\star/\la\bar{q}q\ra)^{1/2}$ from $n=0$ up to
$\sim n=n_0$ (where $n_0$ is nuclear matter density) and then go
over to the linear scaling $m_\rho^\star/m_\rho\sim
\la\bar{q}q\ra^\star/\la\bar{q}q\ra$ up to the chiral transition
density $n_c$ at which the mass is to vanish (in the chiral limit)
according to the vector manifestation fixed point. In the regime
of the linear scaling above $n_0$, the (vector) gauge coupling
constant should fall linearly in $\la\bar{q}q\ra^\star$, vanishing
at the critical point, with the width of the vector meson becoming
steeply narrower. We suggest the in-medium vector meson mass, {\em
both} parametric and pole, which goes smoothly to zero with
increasing density, to be an order parameter for chiral symmetry
restoration. Some issues related to recent RHIC observations are
qualitatively discussed in the appendices. Our main conclusion
that follows from these considerations is that the movement
towards chiral restoration can be reliably described in
Nambu-Jona-Lasinio mean field, with constituent quarks as
variables, although there is probably density discontinuity
following chiral restoration.

\end{titlepage}
\newpage
\renewcommand{\thefootnote}{\arabic{footnote}}
\setcounter{footnote}{0}
\section{Introduction}\label{INTRO}
\setcounter{equation}{0} 
\renewcommand{\theequation}{\mbox{\ref{INTRO}.\arabic{equation}}}
 \itt
In their ``vector manifestation (VM)" of chiral symmetry, Harada
and Yamawaki~\cite{HY:PR} have determined the parameters of their
hidden local symmetry (HLS) by matching to QCD at a sufficiently
high scale $\Lambda\gsim 4\pi f_\pi$ so that the HLS (hadronic)
effective theory overlaps with QCD. When applied to hadrons in
medium, masses and coupling constants in the hadronic theory are
then ``parameteric," dependent {\it intrinsically} on
density/temperature. Renormalization group (RG) equations
determine how they flow as scale is changed. In this note, we show
that the Nambu-Jona-Lasinio (NJL) theory is a caricature theory
with constituent quarks (or quasiquarks) as variables, which
mimics most aspects of the vector manifestation, especially the
parametric behavior of the masses. NJL has the same quadratic
divergence as the HLS, the parametric dependence of the parameters
being run chiefly by the flow of the in-medium condensate $\la
\bar{q}q\ra^\star$. Furthermore the NJL model makes clear the
nonperturbative (solitonic) nature of the vacuum, although the
fixed-point structure of the HLS cannot be constructed until
mesons are ``integrated in." (For a review on NJL, in particular
the bosonization thereof,  see, e.g., Vogel and
Weise~\cite{vogelweise}). The NJL cutoff $\Lambda_{NJL}$ can be
interpreted as the QCD matching scale for constituent quarks.

Introduction of vector mesons immediately raises the QCD matching
scale to $\Lambda\sim 4\pi f_\pi$, the appropriate one for
Wilsonian matching, in a self-consistent way.

Our interpretation of the HLS vector manifestation of chiral
symmetry clarifies: (1) That the use in Brown-Rho (BR)
scaling~\cite{BR91,LPRV} of medium dependent masses is correct,
although the approximately equal scalings such as
$m_\rho^\star/m_\rho\approx f_\pi^\star/f_\pi$ are only correct as
long as RG flow of $f_\pi^\star$ dominates, at low densities or
temperatures, the flow in other variables, especially that of the
(hidden) gauge coupling $g$, becoming important at higher
densities and temperatures.; (2) Once the parameters of the
Lagrangian that are intrinsically dependent on density or
temperature are established, the configuration mixing such as the
Dey-Eletsky-Ioffe mixing~\cite{DEI} of $\rho$ and $a_1$ can be
carried out with these parameters, although it turns out to be
relatively unimportant compared with the BR parametric
effect~\cite{shuryakbrown} with the Dey-Eletsky-Ioffe
low-temperature theorem remaining valid. It then becomes obvious
that the parametric vector ($\rho$) mass should be used in the
Rapp-Wambach (RW) configuration mixing~\cite{rapp} between the
``elementary" $\rho$ (denoted as \rhoelm) and the $\rho$-sobar
$N^* (1520) N^{-1}$ (denoted as $\rho_{sobar}$). Here the
\rhosobar repels the $\rho_{elm}$ and they mix, thus moving the
$\rho_{sobar}$ down in energy and increasing its $\rho$ content,
while pushing up the $\rho_{elm}$. In other words, BR and RW
``fuse" at the lower energies, cooperating in lowering the energy
of the collective excitation, the \rhosobar, which goes to zero
with chiral restoration~\cite{BLRRW}. On the other hand, BR and RW
``defuse", i.e., work in opposite direction for the \rhoelm, where
BR has to overcome RW in order to move the $\rho$-mass down in
energy as found empirically in the STAR collaboration
reconstruction of the $\rho$ from two emitted
pions~\cite{shuryakbrown}. Since BR scaling and RW configuration
mixing give effects of the same general magnitude, it is necessary
to reconstruct their effects in both the region where they $fuse$
and the one in which they $defuse$ in order to disentangle their
effects.

This paper is organized as follows.

In Section 2 we summarize the highlights of the modern
developments.

In Section 3 we briefly review the $conventional$ NJL theory,
pointing out that the cutoff $\Lambda$ is sufficiently high to
consider constituent quarks but only of the order of the
$\rho$-meson mass, so unsuitable for calculating properties of
vector mesons.

In Section 4 we consider the vector manifestation of chiral
symmetry. We show that introduction of $\rho$-meson degrees of
freedom necessitates raising the NJL cutoff by a factor of
$\sqrt{2}$ for consistency in describing the chiral restoration
transition. The factor arises when $\rho$ is present because the
$\rho\pi$ loop correction cancels half of the pion loop which
enters into the running of $f_\pi$ with scale. It follows from the
Wilsonian matching that the flow of the quark condensate $\la
\bar{q}q\ra^\star$ governs the movement of the $\rho$-mass to zero
(in the chiral limit). We shall refer to this NJL implemented with
the VM as ``VM-NJL."

In Section 5 the Rapp/Wambach configuration mixing is argued, when
the parametric structure of the vector manifestation is suitably
implemented, to be an additional and integral -- and not an
alternative -- ingredient to BR scaling.

In Section 6 we shall model the modifications in our dropping mass
scenario for hadronization in RHIC physics. We show that there is
a smooth transition with expansion of the fireball from gluons
into vector and axial-vector mesons, the mesons starting out as
massless and getting their masses back as temperature and density
decrease. We show how the interaction of off-shell vector mesons
can be reconstructed from lattice gauge results for the quark
number susceptibility.

The summary is given in Section 7.

In the appendices are given some qualitative discussions on recent
puzzling observations at RHIC in terms of our vector manifestation
and gluon-vector-meson ``relay" scenario. These sections are quite
preliminary and will have to be quantified.

\section{The Developments}\label{modern}
\setcounter{equation}{0} 
\renewcommand{\theequation}{\mbox{\ref{modern}.\arabic{equation}}}
 \itt
The concept of dropping masses in dense/hot medium formulated in
1991~\cite{BR91} has recently undergone rejuvenation in both
experimental and theoretical sectors. The original idea of
implementing the trace anomaly of QCD in terms of a scalar $\chi$
field which at the mean field level gave rise to the scaling of
dynamically generated hadron masses has recently been
reformulated~\cite{LPRV} in terms of excitations in dense skyrmion
matter, a formulation that renders the notion of BR scaling more
precise and provides its possible link to hidden local symmetry
theory with the vector manifestation (HLS/VM)~\cite{HY:PR}. More
specifically, the result of \cite{LPRV} provides a connection
between the BR scaling and the crucial ``intrinsic" background
dependence brought in by matching to QCD in the HLS/VM theory.
Furthermore, the $\chi$ field embedded in medium, i.e., $\chi^*$,
the average value of which represents the radius of the chiral
circle and shrinks as chiral symmetry is restored is shown by
Shuryak and Brown~\cite{shuryakbrown} to correctly give the
decrease in $\rho$-mass with density in the pristine atmosphere
created in RHIC physics and measured by the STAR collaboration
just preceding freeze-out of the two pions making up the
$\rho$-meson.

This new development is chiefly triggered by a series of
work~\cite{seriesHY} on ``vector manifestation of chiral symmetry"
done by Harada and Yamawaki (reviewed in \cite{HY:PR}). The work
of Harada and Yamawaki is based on the exploitation of hidden
local symmetry (HLS) that emerges via the gauge-equivalence of
nonlinear $G/H$ theory -- where $G$ is the unbroken global
symmetry and $H$ is the subgroup of $G$ that is left unbroken
after spontaneous symmetry breaking $G$ -- to linear $G\times
H_{\rm local}$ theory. This equivalence allows them to extend
chiral dynamics, i.e., chiral perturbation theory, up through the
scale of the vector mesons. We might identify this to be a
``bottom-up" approach starting with a chiral symmetric low-energy
theory. How to go up in scale using the gauge-equivalence argument
is in general not unique; it is known that a variety of gauge
theory Lagrangians can reduce to a non-gauge theoretic effective
Lagrangian, so reversing the reasoning, one arrives at the
conclusion that a given low-energy effective theory, if
unconstrained, can flow into a variety of different directions.
The crux of the observation made by Harada and Yamawaki is that
there is a unique flow if the HLS theory is matched \`a la Wilson
to QCD at a scale in the vicinity of the chiral scale
$\Lambda_\chi$. The matching is made via the vector and
axial-vector correlators between the hadronic sector described by
the effective field theory and the QCD sector described by OPE
with the gluon and quark condensates and color coupling constant
given at the matching scale $\Lambda_M$. The fixed point structure
of the effective field theory then locks in the renormalization
group (RG) flow towards chiral restoration corresponding to the
vector manifestation (VM) fixed point~\footnote{It may be helpful
to specify how the VM of Harada and Yamawaki differs from Georgi's
vector limit (VL)~\cite{georgi}. Georgi's VL is characterized by
$g=0$, $a=1$ and $f_\pi\neq 0$ whereas the VM is given by $g=0$,
$a=1$ and $f_\pi= 0$. What makes the pion decay constant in the VM
flow to zero is the presence of quadratic divergence in the
renormalization-group equation for $f_\pi$ which will figure
importantly in the argument we develop below. Unlike Georgi's
vector limit, there is no symmetry enhancement in the VM at the
chiral restoration which is consistent with the symmetries of QCD.
It should however be mentioned that although perhaps not directly
relevant to QCD, Georgi's symmetry enhancement has a significant
place in the notion of ``theory space" being developed for
dimensional deconstruction in particle physics~\cite{HY:littleH,
SS:littleH}.}.

We have recently suggested~\cite{BR:PR01,BR:BERK,MR:taiwan} for
the three-flavor case that the above argument can be made also in
a ``top-down" theory following Berges-Wetterich's color-flavor
locking mechanism in the Goldstone mode~\cite{wett,berges-wett}.
If one assumes that quark-antiquark pairs condense not only in the
color-singlet channel as in the standard picture but also in the
color-octet channel as was suggested by Berges and Wetterich, then
the resulting effective theory results in a form identical to the
HLS theory of Harada and Yamawaki. Since the variables in this
theory are the massive quasiquarks and Higgsed gluons, it is
properly in QCD language and although there is no proof yet, it is
plausible from the Harada-Yamawaki analysis that the theory
possesses the VM fixed point and hence the same chiral restoration
scenario~\cite{MR:taiwan}.

\begin{figure}[h]
\centerline{\epsfig{file=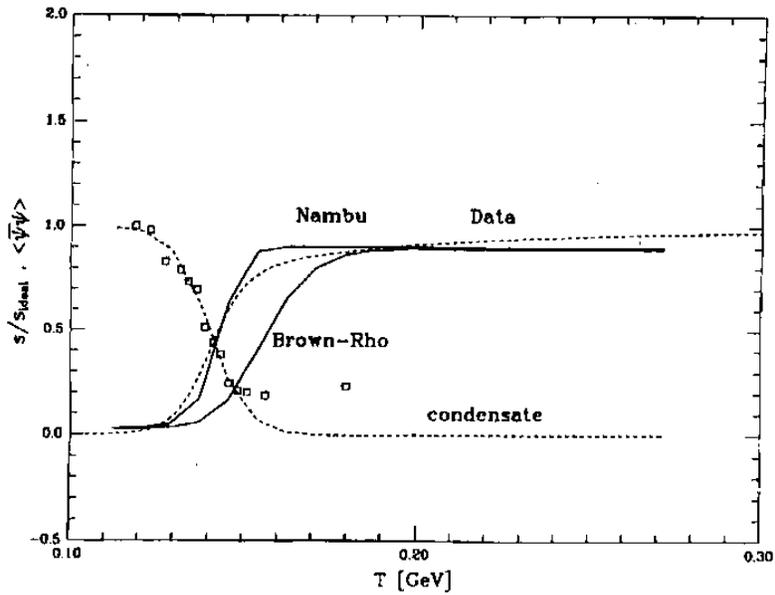,width=9cm,angle=90}}
 \caption{\small Comparison of entropy derived by Koch
and Brown from the lattice data on energy density (dashed line)
and from dropping masses (full lines) calculated in the ``Nambu
scaling" and in the old ``Brown-Rho scaling" which is now
superseded by a revised one described in this paper. In the lower
dashed line for $\la\bar{\psi}\psi\ra$ the bare quark mass has
been taken out. Both $s$ and $\la\bar{\psi}\psi\ra$ given as open
squares were taken from lattice gauge simulations.
}\label{kochbrown}
\end{figure}

The vector manifestation viewed both ``bottom-up" and ``top-down"
implies that the vector-meson mass goes to zero~\footnote{Unless
otherwise noted, we will be mostly considering the chiral limit.}
linearly as $\la \bar{q}q\ra$ does as the critical point is
approached. We will refer to this scaling as ``Nambu scaling." In
fact, in a model calculation connecting quantities calculated
directly by lattice gauge calculations, Koch and
Brown~\cite{kochbrown} found confirmation of Nambu scaling in
temperature. These authors considered the increase in entropy as
24 hadrons went massless with increasing temperature. (This number
of degrees of freedom must be preserved in the transition which
ends up with a quark multiplicity of 24~\cite{BJBP}). With
$\la\bar{q}q\ra^\star$ taken directly from lattice calculations,
the entropy $S$ for assumed scalings
 \be
\frac{M_{Hadron}^\star}{M_{Hadron}}=\left(\frac{\la\bar{q}q\ra^\star}
{\la\bar{q}q\ra}\right)^{1/3}\  {\rm or} \ \
\frac{\la\bar{q}q\ra^\star}{\la\bar{q}q\ra}
 \ee
was compared with the $S$ calculated in the lattice gauge
calculation. We show (for completeness) the comparison in
Fig.\ref{kochbrown}. The rise in lattice gauge entropy through the
chiral restoration transition is well reproduced with the Nambu
scaling.

This linear dependence expected on general ground and observed in
the Koch-Brown analysis has been verified in the framework of HLS
theory only near the critical point both in
temperature~\cite{HS:T} and in density~\cite{HKR}. Recent
works~\cite{HS:Tprog,HKRS} indicate that the interplay of the
``intrinsic temperature/density dependence"~\footnote{This effect
is missing in all low-energy effective theory calculations so far
available in the literature except in ``unified approach" based on
skyrmion soliton model of \cite{LPRV}.} and ``hadronic
thermal/density loop effects" makes the situation considerably
more subtle than naively thought, so except at the two extremes,
i.e., at very low temperature/density and very near chiral
restoration, the scaling behavior is not quantitatively understood
in HLS/VM theory.

It is interesting to note that the strategy involved in the
Harada-Yamawaki approach is related to that used in QCD sum rule
calculations~\cite{adamibrown}. There the hadronic correlators are
also matched with the QCD ones at a scale $\Lambda_{SR} \sim 1
GeV$. A Borel transformation is carried out with the result that
the scale $\Lambda_{SR}$ is replaced by the Borel mass $M_B$.
Thus, terms of the form $\la G_{\mu\nu}G^{\mu\nu}\ra/M_B^4$ and
$\la\bar{q}q\ra^2/M_B^6$ occur in the equations. In parallel to
the Harada-Yamawaki scheme where the bare parameters determined at
the matching scale $\Lambda_M$ are brought down to the on-shell
scale $m_\rho$ (or $m_\rho^*$ in medium) by RGE, the decrease in
scale in the QCD sum rule approach is then accomplished by
bringing the Borel mass down to $m_\rho$ (or $m_\rho^*$), at which
a wide plateau is to be established for the procedure to be
viable. Indeed such a plateau was found by Adami and Brown in
their study of the vector meson mass $m_\rho^*$ in temperature.
They found that as $\la\bar{q}q\ra^*\rightarrow 0$,
$m_\rho^*\rightarrow 0$. Thus the zero in $m_\rho^*$, which is a
fixed point in HLS/VM, was found in the QCD sum rules by changing
only the quark condensate, and leaving the color-gauge coupling
constant $\alpha_s$ and the gluon condensate $\la
G_{\mu\nu}G^{\mu\nu}\ra$ alone. In this formulation, the former
changes perturbatively and the latter makes ``radiative
corrections" to the mass generated by the $\la\bar{q}q\ra$
condensate. A crucial question raised in the QCD sum rule
calculations of Adami and Brown was why $M_B$ should scale with
density and/or temperature and whether the assumed scaling in
order to produce a plateau about $m_\rho^*$ was correct. We
believe this question is answered by HLS/VM theory.

The developments described above clearly indicate that {\it all
chiral effective theories -- regardless of whether they are local
gauge symmetric or not -- should be so constrained in order to
correctly represent QCD}. In this paper, we argue that the
Nambu-Jona-Lasinio (NJL) effective theory can be made equivalent
to the VM fixed point theory by extending it up through the scale
of the vector mesons to the scale at which the hadronic theory can
be matched with QCD and hence can supply the means to interpolate
the scaling between the two extremes $n=0$ and $n=n_c$. To do
this, we formulate NJL with constituent quarks (or quasiquarks) as
variables. We assume that before chiral restoration, transition
from nucleons to quasiquarks can be
made~\cite{BBR96,pirner}~\footnote{This must be a good assumption
as we know the quasiquark picture is fairly good even in free
space.}. The phase transition is most naturally made in these
variables with continuity in the number of degrees of freedom.

\section{Nambu-Jona-Lasinio Theory}\label{NJL}
\setcounter{equation}{0} 
\renewcommand{\theequation}{\mbox{\ref{NJL}.\arabic{equation}}}
\itt The NJL has the same quadratic divergence in its gap
equation, the equation governing the dynamical generation of mass,
as the VM of Harada and Yamawaki~\footnote{We are using the cutoff
to regulate the theory. As discussed by Harada and Yamawaki in
\cite{HY:PR}, the use of cutoff does not spoil chiral symmetry in
the gap equation while it does in the correlators.}. We wish to
study chiefly the NJL $\la\bar{q}q\ra$ condensate, since its
medium dependence is found (as we discuss below) to run the
dropping of the $\rho$ meson mass.

Following the work of Hansson et al~\cite{hansson}, we note that
in the vacuum, NJL theory gives
 \be
\la\bar{q}q\ra=-\frac{N_f N_c m_Q}{4\pi^2}\left[\Lambda_4^2
-{m_Q}^2\ln\left(\frac{\Lambda_4^2+{m_Q}^2}
{{m_Q}^2}\right)\right] \label{qbarq}
 \ee
where $\Lambda_4$ is the covariant cutoff and $m_Q$ is the
dynamically generated quark mass of the quasiquark. If one takes
$\Lambda_4=715$ MeV, one obtains, for $N_f=2$ and $N_c=3$,
 \be
\la\bar{q}q\ra=-(250\ {\rm MeV})^3.
 \ee
How the condensate changes in medium in this model has been worked
out simply and elegantly by Bernard, Meissner and
Zahed~\cite{bernard} who found~\footnote{In the chiral limit their
critical density was $\sim 2n_0$, much too low, but with bare
quark mass $\bar{m}=(m_u+m_d)/2\approx 5$ MeV it was more like the
$5n_0$ that we quote here. See the discussion at the end of
Section 4.} a critical density of $\sim 5 n_0$ (where $n_0$ is the
normal nuclear matter density) and $T_c\gsim 200$ MeV with a
cutoff of 700 MeV. We believe this to be ballpark also for the
vector manifestation which we discuss in the next section.

In adding nucleons in positive energy states, each added nucleon
brings in the same scalar condensate as the negative energy
nucleon contributes -- but negatively -- to $\la\bar{q}q\ra$ in
eq.(\ref{qbarq}). Thus we get
 \be
\la\bar{q}q\ra^\star&=& \la\bar{q}q\ra +\frac{6
m_Q}{4\pi^2}\left[k_F^2 -{m_Q}^2\ln\left(\frac{k_F^2+m_Q^2}
{{m_Q}^2}\right)\right]\nonumber\\
&=& -\frac{6 m_Q}{4\pi^2}\left[\Lambda_4^2-k_F^2
-{m_Q}^2\ln\left(\frac{\Lambda_4^2+{m_Q}^2}
{k_F^2+{m_Q}^2}\right)\right]. \label{qbarqstar}
 \ee
Note that when the Fermi momentum $k_F$ reaches the cutoff
$\Lambda_4$, the condensate vanishes. An intuitive interpretation
is to view the cutoff as the ``Fermi momentum" of the sea of the
condensed negative energy quarks and that it reflects the
magnitude of $\la\bar{q}q\ra$. The condensate vanishes as the
positive energy sea becomes as large in magnitude as the negative
energy sea characterized by $\Lambda_4$. This would imply a smooth
mean field transition with the vector mass going to zero with the
order parameter. Since the behavior of the vector degrees of
freedom signals the phase transition, it is likely that there is a
condensation of these degrees of freedom at the critical density
$n_c$; i.e., a discontinuity in density, the $\omega$-meson being
the easiest to condense~\cite{LR}.
We note that $\la\bar{q}q\ra$ scales as
 \be
\la\bar{q}q\ra\propto m_Q\Lambda_4^2\label{qbarqscale}
 \ee
for $\Lambda_4\gg m_Q, \ k_F$, which is the situation up through
nuclear matter density $n_0$.

We shall take the usual position on NJL and assume $\Lambda_4$ to
be constant, independent of density or temperature. This makes it
possible to make a connection to the vector manifestation of
chiral symmetry as we discuss in the next section.

Given $\Lambda_4$ constant, eq.(\ref{qbarqscale}) shows that in
going to finite density, we expect, at low density, in the mean
field approximation
 \be
m_Q^\star\propto \la\bar{q}q\ra^\star\propto f_\pi^\star\label{mQ}
 \ee
with the star representing in-medium quantities. In going from
(\ref{qbarqscale}) to (\ref{mQ}), we have used the fact that $m_Q$
in medium satisfies an in-medium gap equation and hence is density
dependent. The last relation of (\ref{mQ}) follows from the
Goldberger-Treiman relation that is expected to hold in medium at
mean field in NJL as~\footnote{We are fully aware of the fact that
since Lorentz invariance is broken in medium, one has to
distinguish the time and space components of $f_\pi$ here as well
as in what follows. We are keeping Lorentz invariant notation but
as discussed in \cite{HKR,LPRV}, the main argument holds valid:
This technical matter can be easily attended to.}
 \be
G_{\sigma QQ}m_Q^\star=f_\pi^\star.
 \ee
We have used that the coupling $G_{\sigma QQ}$ does not scale at
mean field. We shall refer to the scaling (\ref{mQ}) as ``Nambu
scaling"\cite{BR:PR01}.

In \cite{BR:PR96} following the argument of Lutz et
al~\cite{lutzetal}, it was outlined that from the
Gell-Mann-Oakes-Renner (GMOR) relation
 \be
f_\pi^2 m_\pi^2=-\bar{m}\la\bar{q}q\ra\label{GMOR}
 \ee
where $\bar{m}$ is the bare quark mass which does not scale with
density,
 \be
{f_\pi^\star}^2=-\bar{m}\frac{\la\bar{q}q\ra^\star}{{m_\pi^\star}^2}.\label{GMRO1}
 \ee


Although there have been many papers on how the ``pion
mass"~\footnote{We are being a bit cavalier in what we mean by
``pion mass." To be precise, one has to indicate whether one is
talking about the ``parametric mass" or pole mass. Although to the
order we consider, there is no difference between the two, we have
in mind the parametric mass and not the pole mass.} behaves in
medium, it is a very subtle issue as shown in \cite{LPMRV} as the
in-medium Goldstone boson mass is subject to intricate
self-consistency conditions that non-Goldstone bosons are not
subject to. Given this situation, we take the point of view --
supported by \cite{LPMRV} -- that the (parametric) pion mass that
appears in current algebra relations is protected (at least) at
low density by chiral symmetry in which case we would have the
scaling
 \be
f_\pi^\star\propto (\la\bar{q}q\ra^\star)^{1/2}\label{scalelow}
 \ee
that differs from the Nambu scaling (\ref{mQ}). Empirically, this
is much more reasonable at low density than the Nambu scaling.
This implies that
 \be
m_\rho^\star\sim f_\pi^\star g^\star \sim f_\pi^\star\sim
{\la\bar{q}q\ra^\star}^{1/2}
 \ee
where we have used that the (hidden) gauge coupling does not scale
appreciably for $n<n_0$.

On the other hand, Harada, Kim and Rho find that the vector
manifestation implies {\em close} to the chiral restoration
density/temperature,
 \be
\frac{m_\rho^\star}{m_\rho}\sim
\frac{\la\bar{q}q\ra^\star}{\la\bar{q}q\ra}\label{linear}
 \ee
in agreement with Koch and Brown~\cite{kochbrown}. What is
surprising is that the mass scaling seems to accelerate to the
form (\ref{linear}) at densities only slightly higher than $n_0$.
As we shall see in the next section, $m_\rho^\star$ goes to zero
as $f_\pi^\star g^\star \sqrt{a}$, where both $f_\pi^\star$ and
$g^\star$ go to zero and $a$ flows towards its fixed point $a=1$.
In fitting to various theories of nuclear matter at densities
$n\gsim n_0$, Song et al~\cite{songetal,song} arrived at that the
vector (gauge) coupling $g$ decreases such that the vector mean
field which goes as ${g^\star}^2/{m_V^\star}^2$ was roughly
constant above $n_0$, although the $m_V^\star\rightarrow 0$
evolution is dictated by the flow towards the fixed point. This
observation combined with that given above suggests that the old
BR scaling
 \be
\frac{f_\pi^\star}{f_\pi}\approx \frac{m_\rho^\star}{m_\rho}
 \ee
to hold only up to densities where $g^\star$ begins to decrease
appreciably, $\frac{m_\rho^\star}{m_\rho}$ dropping from then on
faster than $\frac{f_\pi^\star}{f_\pi}$. For the higher densities
$n\gsim n_0$, where $g^\star$ scales roughly as $m_\rho^\star$, we
find from the flow towards the fixed point that
 \be
\frac{m_\rho^\star}{m_\rho}\sim
\left(\frac{f_\pi^\star}{f_\pi}\right)^2\sim
\frac{\la\bar{q}q\ra^\star}{\la\bar{q}q\ra}.
 \ee
Although this iterative scheme fails if carried to higher order,
it does indicate that the more rapid Nambu scaling sets in above
$n_0$.

From the linear decrease in $\la\bar{q}q\ra^\star$ with
density~\cite{levin,cohen},
 \be
\frac{\la\bar{q}q\ra^\star}{\la\bar{q}q\ra}\simeq 1-
\frac{\Sigma_{\pi N} n}{f_\pi^2 m_\pi^2}
 \ee
one finds with $\sigma_{\pi N}=46$ MeV that
 \be
\frac{\la\bar{q}q\ra^\star}{\la\bar{q}q\ra}\simeq 0.64
 \ee
at nuclear matter density, giving
 \be
\frac{f_\pi^\star}{f_\pi}\simeq 0.8.\label{fpi}
 \ee
We should note that the effective mass of the nucleon
$m_N^\star/m_N$ drops faster, by a factor $\sqrt{g_A^\star/g_A}$,
which is a loop correction going beyond the mean
field~\cite{BR91}. (\ref{fpi}) is consistent with the scaling
$m_V^\star/m_V (n_0)=\Phi (n_0)\approx 0.78$ found in the analysis
of nuclear matter based on the mapping of Landau Fermi-liquid
theory to an effective chiral field theory with parameters running
\`a la BR scaling~\cite{songetal,FRS,song,MR:MIGDAL}. From our
above argument, at densities $n\gsim n_0$, Nambu scaling sets in,
so $(f_\pi^\star/f_\pi)^2$ decreases roughly linearly with density
from then on, going to zero at $\sim 5n_0$. Note that this is
roughly the density of chiral restoration in NJL quoted above.
Thus, the scaling of masses with Nambu scaling above $n_0$ is
consistent with the $n_c$ obtained in NJL.

Amusingly, the consequence of this accelerated scaling is that the
vector meson mass decreases roughly linearly with density $n$, all
the way from $n=0$ to $n_c\approx 5 n_0$.


As $n$ nears $n_c$, eq.(\ref{qbarqstar}) takes the form
 \be
\la\bar{q}q\ra^\star\propto m_Q^\star \Lambda^\star\Lambda_4
 \ee
where $\Lambda^\star=\Lambda_4-k_F$. One can think of
$\Lambda^\star$ as an $effective$ cutoff measured relative to the
Fermi surface that goes to zero as the condensate goes to
zero~\footnote{We are grateful to Youngman Kim for pointing out
this observation.}. It is actually a cutoff in the phase space
available to a positive energy quark before chiral restoration.

In summary, NJL gives a simple picture of chiral restoration. In
mean field, the $\la\bar{q}q\ra^\star$ condensate goes smoothly to
zero, as positive energy quarks are added. Hadron masses follow
this behavior of the condensate near chiral restoration. This does
not mean, however, that the phase transition is mean field. There
probably is a density discontinuity as suggested by Langfeld and
Rho due to an induced symmetry breaking of vector
currents~\cite{LR}.

\section{The Vector Manifestation of Chiral Symmetry}\label{VM}
\setcounter{equation}{0} 
\renewcommand{\theequation}{\mbox{\ref{VM}.\arabic{equation}}}
\itt Weinberg realized a long time ago that chiral symmetry could
be extended to straightforwardly include the $\rho$-meson in
phenomenological Lagrangians~\cite{weinbergvector}. This idea was
put into a local gauge-invariant form by Bando et
al~\cite{bandoetal} and summarized by Bando, Kugo and
Yamawaki~\cite{BKY-PR}, in their formulation of the nonlinear
sigma model as a linearized hidden local symmetry (HLS) theory in
which the $\rho$-meson is the gauge particle. Since the vector
degrees of freedom are strongly coupled to the nucleon, these must
be included in the hadronic sector up to a scale of
$\Lambda_\chi\sim 4\pi f_\pi$ where they can be matched with QCD.
The quadratic divergence in Wilsonian renormalization group
equations of HLS which was recently recognized by Harada and
Yamawaki changed the structure of the theory radically. In
particular it contributes a quadratic term $\Lambda^2$ in the RG
flow of the order parameter $f_\pi$ and, therefore, is of primary
importance in the chiral restoration transition.~\footnote{Note
that power divergences do not appear in chiral perturbation theory
in the regime where the chiral expansion is valid. However
whenever one confronts anomalous scales as when one has
quasi-bound states or phase transitions, power divergences play
the key role. This is the case here in the vector manifestation
where chiral restoration occurs. Another case that has been
discussed in a different context is the linear divergence that
figures in low-energy nucleon-nucleon scattering in S-wave that
has a scale invariant fixed point corresponding to an infinite
scattering length. See \cite{MR:taiwan} for discussions on this
matter and references.}

In a spirit similar to QCD sum rules, a Wilsonian matching is
carried out at a scale $\Lambda \gsim 4\pi f_\pi$. This matching
determines the $bare$ parameters of the HLS Lagrangian. The scale
is then systematically brought down all the way to zero by the
renormalization group equations developed in the effective
hadronic sector. The loop corrections are carried out with the
$\rho$-mass counted of the same chiral order as the pion mass. As
first pointed out by Georgi~\cite{georgi}, this is the only way
that the $\rho$ field can be incorporated in a systematic chiral
perturbation theory. Considering the $\rho$-mass as ``light" seems
unjustified in free space but it makes sense when the $\rho$-mass
drops in medium as mentioned above and developed further below.
Even in the vacuum, this counting rule combined with the Wilsonian
matching at $\Lambda\gsim 4\pi f_\pi$ is found to reproduce chiral
perturbation theory \`a la Gasser and Leutwyler~\cite{chpt} in
$\pi\pi$ interactions up to $\calO (p^4)$.

The Harada-Yamawaki work is summarized in their Physics
Reports~\cite{HY:PR}. We briefly review the parts relevant to our
present work.

Harada and Yamawaki matched the hadronic vector-vector and
axial-vector-axial-vector correlators defined by
 \be
i\int d^4x e^{iqx}\la 0|TJ_\mu^a (x)J_\nu^b (0)|0\ra
&=&\delta^{ab}
\left(q_\mu q_\nu -g_{\mu\nu} q^2\right) \Pi_V (Q^2),\nonumber\\
i\int d^4x e^{iqx}\la 0|TJ_{5\mu}^a (x)J_{5\nu}^b (0)|0\ra
&=&\delta^{ab} \left(q_\mu q_\nu -g_{\mu\nu} q^2\right) \Pi_A
(Q^2)
 \ee
with
 \be
Q^2\equiv -q^2.
 \ee
at the scale $\Lambda \gsim 4\pi f_\pi$ between the hadronic
sector and the QCD sector. In the hadronic sector, they are given
by the tree terms~\footnote{Since the correlators are computed at
the matching scale, there are no loop corrections.}
 \be
\Pi^{(HLS)}_A (Q^2)&=&\frac{F_\pi^2 (\Lambda)}{Q^2}-2 z_2
(\Lambda),\\
\Pi^{(HLS)}_V (Q^2)&=&\frac{F_\sigma^2 (\Lambda)}{M_v^2
(\Lambda)+Q^2}\left[1-2g^2(\Lambda)z_3 (\Lambda)\right]-2z_1
(\Lambda)
 \ee
with
 \be
 M_v^2 (\Lambda)\equiv g^2(\Lambda)F_\sigma^2 (\Lambda).
 \ee
Here $g$ is the HLS coupling and $z_i$'s are $\calO (p^4)$ terms
in the Lagrangian. In the quark-gluon sector, the correlators are
given by the OPE to $\calO (1/Q^6)$,
 \be
\Pi^{(QCD)}_A&=&\frac{1}{8\pi^2}\left[-(1+\alpha_s/\pi)\ln\frac{Q^2}{\mu^2}
+\frac{\pi^2}{3}\frac{\la\frac{\alpha_s}{\pi}G_{\mu\nu}
G^{\mu\nu}\ra}{Q^4}
+\frac{\pi^3}{3}\frac{1408}{27}\frac{\alpha_s\la\bar{q}q\ra^2}{Q^6}\right],\\
\Pi^{(QCD)}_V&=&\frac{1}{8\pi^2}\left[-(1+\alpha_s/\pi)\ln\frac{Q^2}{\mu^2}
+\frac{\pi^2}{3}\frac{\la\frac{\alpha_s}{\pi}G_{\mu\nu}
G^{\mu\nu}\ra}{Q^4}
-\frac{\pi^3}{3}\frac{896}{27}\frac{\alpha_s\la\bar{q}q\ra^2}{Q^6}\right]
 \ee
where $\mu$ here is the renormalization scale of QCD (e.g., in
dimensional regularization). Chiral restoration is associated with
$\la \bar{q}q\ra^\star$ going to zero. In this case
$\Pi_A^{(QCD)}$ and $\Pi_V^{(QCD)}$ are equal {\it for all} $Q^2$.

The Wilsonian matching equates the correlators of QCD with those
of HLS and also their derivatives. One obtains
 \be
\frac{F_\pi^2(\Lambda)}{\Lambda^2}=-Q^2\frac{d}{dQ^2}\Pi_A^{(QCD)}
(Q^2)|_{Q^2=\Lambda^2}.
 \ee
The quark and gluon condensates are obtained phenomenologically as
in the QCD sum rules, and $\alpha_s (\Lambda)$ from the
perturbative expression.

The Wilsonian matching implies that
 \be
\frac{F_\pi^2 (\Lambda)}{\Lambda^2}
=\frac{1}{8\pi^2}\left[1+\frac{\alpha_s}{\pi}+\frac{2\pi^2}{3}
\frac{\la\frac{\alpha_s}{\pi}G_{\mu\nu}
G^{\mu\nu}\ra}{\Lambda^4}\right]\neq 0
 \ee
which implies in turn that matching with QCD dictates
 \be
f_\pi^2 (\Lambda)=F_\pi (\Lambda)\neq 0
 \ee
even though at the point of chiral restoration
 \be
f_\pi^2 (0)=0.
 \ee
Note that $F_\pi (\Lambda)$ is a parameter of the Lagrangian
defined at the scale $\Lambda$ and $f_\pi (\Lambda)$ is the pion
decay constant at the same scale. Thus, $f_\pi (\Lambda)=F_\pi
(\Lambda)$ is not the order parameter but just a parameter of the
$bare$ HLS Lagrangian defined at the cutoff $\Lambda$ where the
matching to QCD is made. At chiral restoration (at $n=n_c$ or
$T=T_c$) characterized by
$\la\bar{q}q(\Lambda;n_c)\ra^\star=\la\bar{q}q(\Lambda;T_c)\ra^\star
=0$, one finds
 \be
g(\Lambda)&\rightarrow& 0,\nonumber\\
a(\Lambda)=F^2_\sigma (\Lambda)/F^2_\pi (\Lambda)&\rightarrow&
1,\nonumber\\
z_1 (\Lambda)-z_2 (\Lambda)&\rightarrow& 0,\nonumber\\
F_\pi (\Lambda)&\neq& 0.
 \ee
Here $F_\sigma$ is the decay constant for the scalar Goldstone
corresponding to the would-be longitudinal component of the
massive $\rho$-meson in the Nambu-Goldstone phase.

The HLS theory has an elegant fixed-point structure, dictated by
the renormalization group equations (RGEs). As in Georgi's vector
limit, one has the fixed points
 \be
g&=&0,\nonumber\\
a&=&1.
 \ee
Below the $\rho$-meson mass the RGE for $f_\pi$ is the same as in
the nonlinear chiral Lagrangian
 \be
\mu\frac{d}{d\mu} f^2_\pi (\mu)=\frac {2N_f}{(4\pi)^2}
\mu^2\label{fpiRG}
 \ee
where $N_f$ is the number of flavors and $\mu$ the renormalization
point. The rapid scaling with $\mu^2$ near the phase transition
comes from the quadratic divergences. Chiral restoration occurs
for
 \be
f_\pi (0)=0
 \ee
at the scale corresponding to pion on-shell. If we integrate
(\ref{fpiRG}) up to the scale $\bar{\Lambda}$, say, at which the
vector meson is integrated out, then
 \be
f_\pi^2=\frac{N_f}{(4\pi)^2} \bar{\Lambda}^2
 \ee
which gives
 \be
\bar{\Lambda}=\frac{4\pi}{\sqrt{2}} f_\pi (\bar{\Lambda}).
 \ee
If we now adopt at this point the KSRF relation $m_\rho^2=2
f_\pi^2 g_V^2$, then $f_\pi$ (to be evaluated at the pion
on-shell) is 93 MeV (88 MeV in the chiral limit). Thus
 \be
\bar{\Lambda}=827 (782)\ {\rm MeV}.
 \ee
This is just the cutoff employed in conventional NJL calculations,
so everything looks fine, except that it is too low a scale to
match to QCD and have a description of $\rho$-meson properties.

Now once the $\rho$-meson is included in the HLS, the $(\rho,
\pi)$-loop in the axial-vector-axial-vector correlator -- which is
used to calculate $f_\pi$ -- cancels half of the pion loop with
the result that in this theory~\footnote{In order for this
cancellation to take place, the scale must be above the
$\rho$-meson, which is assumed in our consideration to be near
zero; in other words, the argument here is made in the vicinity of
chiral restoration.}
 \be
\Lambda =4\pi f_\pi (\Lambda).
 \ee
The NJL with the extended cutoff gives the VM-NJL. Not only is
$\Lambda$ a factor of $\sqrt{2}$ greater than $\bar{\Lambda}$ but
$f_\pi$ at this higher scale is greater than $f_\pi$ at $m_\rho$.
Consequently, $\Lambda
>1$ GeV is easily employed as the matching scale. Thus, the
introduction of the $\rho$-meson in the loop is necessary to have
the cutoff evolve to a sufficiently high scale in order to do the
Wilsonian matching.

We should stress that the linear dependence of $f_\pi$ on the
cutoff $\Lambda$ was already contained in NJL, indicating again
the relevance of NJL to the Harada-Yamawaki theory. The point here
is that the cutoff in NJL can be interpreted as effectively
playing the role of Wilsonian matching scale, as noted before, and
the H-Y theory instructs us how to extend NJL through $m_\rho$ to
$4\pi f_\pi$.

Since we have the Wilsonian matching, we find that the
$(\la\bar{q}q\ra^\star)^2$ terms in $\Pi_A^{(QCD)}$ and
$\Pi_V^{(QCD)}$ completely dominate the right-hand side, the flow
to the fixed point locking their flow toward zero.
The other terms in the correlator are down by an order of
magnitude. Thus, the flow in $\la\bar{q}q\ra^\star$ can be
obtained from NJL provided the parameters, in particular, the
cutoff $\Lambda_4$, are properly chosen.

\begin{figure}[hbt]
\vskip 0cm
 \centerline{\epsfig{file=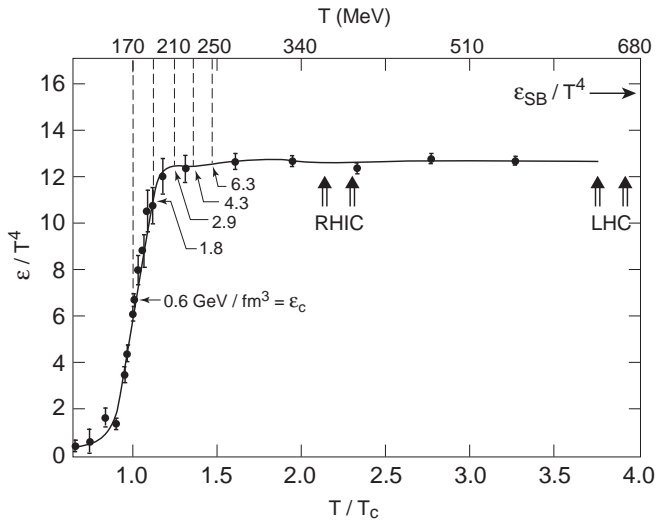,width=10cm,angle=0}}
\vskip 0cm \caption{\small Energy density in units of $T^4$ for
QCD with 3 dynamical light quark
flavors~\cite{karsch,heinz-karsch}.}\label{heinz}
\end{figure}

Our strategy is to choose the NJL parameters in order to model
results of lattice gauge calculations for finite temperature, and
then use the NJL results at finite density to give the flow to the
vector manifestation (VM) fixed point. Although this will be a
continuing activity to obtain improved NJL parameters as lattice
gauge simulations (LGS) are improved, we note that it is already
possible to close in on such an analysis. The NJL calculation of
Bernard et al~\cite{bernard} with cutoff $\Lambda_4=700$ MeV, not
far from 715 MeV, gives a $T_c\gsim 200$ MeV, essentially that of
the LGS of Karsch~\cite{karsch} which is shown in Fig.\ref{heinz}.
($T_c$ in our scheme is unambiguously defined in the chiral limit
as the temperature at which the hadron masses are zero.) We
require that $\sqrt{2}\Lambda_4\gsim 1$ GeV so that the Wilsonian
matching can be carried out in VM-NJL.

We now shortly discuss how the VM-NJL-LGS link can be extended to
finite densities $n$. For the finite density the explicit chiral
symmetry breaking seems important in NJL and hence must be taken
into account there, because the effects from $\bar{m}$ are
magnified by the condensate $\la\bar{q}q\ra^\star$ in the
Gell-Mann-Oakes-Renner equation (\ref{GMRO1}), the
$\la\bar{q}q\ra^\star$ extending well beyond the $n_c
(\bar{m}=0)$. In Bernard et al with $\bar{m}=5$ MeV, we see that
$\la\bar{q}q\ra^\star$ extends up to $n_c\sim 5n_0$; i.e.,
 \be
n_c (\bar{m}=5 \ {\rm MeV})\simeq 5n_0.
 \ee
We see that even though small bare quark masses enter, they have a
major effect on the phase transition. This is also true with
temperature, where the bare quark masses convert the chiral
transition into a smooth crossover. It is thus useful to have
VM-NJL since the bare quark masses used in the LGS can be put into
NJL.

\section{Back to Brown/Rho Plus Rapp/Wambach}\label{vs}
\setcounter{equation}{0} 
\renewcommand{\theequation}{\mbox{\ref{vs}.\arabic{equation}}}
\itt Now that we have established that the $\rho$-meson mass must
scale with the quark condensate, we need to revisit the Brown/Rho
vs. Rapp/Wambach debate in connection with the CERES dilepton
data. We will propose that the correct approach is ``Brown/Rho +
Rapp/Wambach" with the latter suitably modified to implement the
vector manifestation.

We should recall that the key information from the vector
manifestation is that when matched \`a la Wilson to QCD, the bare
HLS Lagrangian carries parameters that have intrinsic density
dependence. Brown-Rho (BR) scaling~\cite{BR91} exploits this
density dependence in formulating it at the constituent quark
level~\cite{MR:taiwan}. On the other hand, the Rapp/Wambach
approach~\cite{rapp} is formulated in the language of nucleons and
isobars; i.e., in the conventional language of nuclear
interactions. As such, the Rapp/Wambach is immediately accessible
through standard nuclear physics with configuration mixing. The
matching with the experiments is done typically at tree level with
phenomenological Lagrangians. Now this is of course a
well-justified phenomenological approach in very low density
regimes. But the question arises how this approach can access a
higher density regime without the account of the intrinsic density
dependence in the parameters of the Lagrangian. In \cite{BLRRW},
Brown, Li, Rapp, Rho and Wambach suggest how to go from
Rapp/Wambach to Brown/Rho by endowing the vector meson mass {\em
parameter} in Rapp/Wambach with a suitable density dependence. We
reexamine this issue in the light of the vector manifestation.

Rapp et al~\cite{rappetal} use a phenomenological Lagrangian of
the form
 \be
\calL^{S-wave}_{\rho N N^\star}= \frac{f_{\rho N
N^\star}}{m_\rho}\Psi_{N^\star}^\dagger(q_0
\vec{s}\cdot\vec{\rho}_a  -\rho_a^0 \vec{s}\cdot\vec{q})t_a\Psi_N
+{\rm h.c.}
 \ee
to couple the nucleon to the $N^\star (1520)$ resonance in the
$\rho$ channel. They use~\footnote{They actually use $f_{\rho
N^\star N}= 9.2$ but their form factor cuts it down to 7 for the
relative kinematics.}
 \be
f_{\rho N^\star N}\approx 7
 \ee
from fits to the photoproduction data. The $N^\star$ and $N^{-1}$
are combined to have the quantum number of the $\rho$-meson, the
``$\rho$-sobar" denoted \rhosobar. The two branches in the $\rho$
spectral function can be located by solving the real part of the
$\rho$-meson dispersion relation (at $\vec{q}=0$)
 \be
q_0^2=m_\rho^2 + {\rm Re}\Sigma_{\rho N^\star N}
(q_0).\label{disp}
 \ee
Introducing also the backward-going graph, the $N^\star (1520)
N^{-1}$ excitation contributes to the self-energy at nuclear
matter density $n_0$ as
 \be
\Sigma_{\rho N^\star N} (q_0)=\frac 83 f_{\rho N N^\star}^2
\frac{q_0^2}{m_\rho^2} \frac {n_0}{4} \left(\frac{(\Delta
E)^2}{(q_0+i\Gamma_{tot}/2)^2-(\Delta E)^2}\right)
 \ee
where $\Delta E=1520-940=580$ MeV and
$\Gamma_{tot}=\Gamma_0+\Gamma_{med}$ is the total width of the
$N^\star (1520)$ in medium.  When in-medium corrections to the
width are ignored, there are two solutions to eq.(\ref{disp})
 \be
q_0^-\simeq 540\ {\rm MeV}, \ \ q_0^+\simeq 860\ {\rm MeV}.
 \ee
With inclusion of the widths, both resonances, especially the
lower one, are substantially broadened. It is the strength in this
lower resonance which, because of larger Boltzmann factor, gives
rise to the excess dileptons at low invariant masses.

From the Harada and Yamawaki work we learn that the $m_\rho$ in
the preceding expressions should be replaced by $m_\rho^\star$,
essentially the BR scaling mass~\footnote{More precisely the
intrinsic-density-dependent mass $parameter$ of the HLS Lagrangian
$M_\rho (\Lambda; n)$ in \cite{HKR}.}. This would substantially
increase the strength in the lower branch of the $\rho$ and move
it down in energy~\footnote{Note that this VM mechanism provides
the justification for replacing $m_\rho$ by $m_\rho^\star$ (or
more precisely the intrinsic mass $M_\rho^\star$) which was
conjectured in \cite{BLRRW}.}.

There is considerable disagreement in the literature about the
strength of coupling of the $\rho$ to the \rhosobar. Lutz et
al~\cite{lutzetal} obtain essentially the same value as was
obtained by Riska and Brown~\cite{riskabrown} in the chiral quark
model $f_{\rho N^\star N}\simeq 1.44$. However
Riska~\footnote{Private communication from D.O. Riska.} has
pointed out that substantial two-quark contributions have been
left out in the latter work. The work of Nacher et
al~\cite{nacheretal} on the other hand has $f_{\rho N^\star
N}\approx 6$, with inclusion of form factor for the relative
kinematics. Oset suggests in private communication that a little
stronger $\rho$ coupling would be even preferred. The work of
Nacher et al was used to determine how much of the two-pion decay
in the $\gamma p\rightarrow \pi^+\pi^0 n$ process goes through the
$\rho$ content of the $N^* (1520)$ excitation in the Langg\"artner
et al~\cite{langg} experiment. The latter authors arrive at a 20\%
content, in the middle of the 15-25\% quoted in the Particle Data
Physics Booklet. These limits were arrived at from the K-matrix
analysis of Longacre and Dolbeau~\cite{long} and other works
quoted in the booklet. Finally, Huber et al~\cite{huber} find a
photoproduction cross section for the $\rho^0$ of $10.4\pm 2.5\
\mu b$ in the three $^3$He nucleons in the 620-700 MeV range of
photon energy, relevant for the \rhosobar, whereas Langg\"artner
et al find $\sim 3.5\ \mu b$/nucleon for $\rho^0$ production on a
proton. In R/W the production per nucleon in $^3$He should be
higher than on a single nucleon; fused with B/R it should be
nearly double. Helicity measurements enable the Regina group
(Huber, Lolos et al) to separate out the $\rho^0$ contribution
unambiguously and these should remove the possible factor of 2
uncertainty in the cross section, assuming our theory to be
correct.

It is quite natural that R/W describes most of the ``$\rho$"
participation in lower energy nuclear physics. The \rhosobar\
comes at zero density  at 580 MeV, $\sim 3/4$ of the elementary
$\rho$ mass. It would take a density of $\sim\frac 54 n_0$ for BR
scaling to bring the elementary $\rho$ that low. With increasing
density more and more $\rho$ is mixed into the \rhosobar.
Quantitatively, the amount is increased substantially by the
fusing of B/R with R/W. However R/W produces too many
$\rho$-mesons in the region above and about the energy of the
elementary $\rho$, and B/R must be introduced to bring these
masses down.

\section{Hadronization at RHIC}\label{RHIC}
\setcounter{equation}{0} 
\renewcommand{\theequation}{\mbox{\ref{RHIC}.\arabic{equation}}}
\itt Based on the Harada-Yamawaki scenario we define $T_c$ as the
temperature at which the vector mesons go massless in the chiral
limit. In order to talk about temperature we assume equilibration.
Hydodynamical descriptions which imply equilibration seem to set
in fairly early at times $\tau\lsim 1$ fm/c at
RHIC~\cite{kolbetal}.

As discussed by Brown, Buballa and Rho~\cite{BBR96}, results of
lattice gauge simulations indicate that in going upwards through
the chiral restoration about half of the trace anomaly, amounting
to $\sim 250$ MeV/fm$^3$, disappears~\footnote{These authors
attribute this to the Walecka scalar field energy.}. This suggests
that in going downwards through the transition this energy is
available to put the hadrons back on shell. The feeding back of
this energy tends to keep the temperature constant, mimicking the
mixed phase usually assumed in the phase transition. Recall that
the masses change with $\la\bar{q}q\ra^*$ near $T_c$, rather than
with $\sqrt{\la\bar{q}q\ra^*}$ relevant at low $T$. It may take a
relatively long time for the particles to get their masses back;
i.e., to go on shell. In this sense, the phase in which the
hadrons go back on shell is similar~\footnote{We can assess the
accuracy of the assumption of constant $T$ in the following way:
From the calculation of \cite{bernard} and our argument of Section
3 we see that the particles have about half of their mass back
when the temperature has dropped 5\% from $T_c$.} to the long
mixed phase~\cite{BBR96}. This phase may not be, strictly
speaking, the mixed phase associated with the first-order phase
transition but we shall loosely refer to this as such.

One of the remarkable results from RHIC physics from RHIC
physics~\cite{peter-johanna} is the common freeze-out temperature
of $T_{fo}=174\pm 7$ MeV for $all$ hadrons. For a $\rho$ meson of
mass $m_\rho=770$ MeV this means a total energy of $\sim 1090$
MeV.

As developed in \cite{BJBP} quasi-equilibration between $\pi$,
$\rho$ and $a_1$ is probably reached early in the ``mixed phase,"
before the particles have gone all the way on shell. The $\rho$
continues to equilibrate with two pions $\rho\leftrightarrow 2\pi$
long after the mixed phase, essentially down to thermal
freeze-out. In fact, because of its large width, Boltzmann factors
on the two pions distort its shape, essentially lowering its
central value by $\sim 23$ MeV in energy and another lowering of
$\sim 40$ MeV comes because it freezes out slightly off-shell (due
to BR scaling). Because of the above special feature of the
$\rho$, Braun-Munzinger et al~\cite{peter-johanna} do not include
it in their Table. However, we believe that the
quasi-equilibration of the $\pi$, $\rho$ and $a_1$ forms an
underlying framework for the equilibration of the less strongly
interacting particles. For simplicity let us assume that they all
freeze out with a mass of $\sim 95\%$ of their free space mass in
the mixed phase. Then a drop in $T_{fo}$ of 5 \% will correct the
Braun-Munzinger et all temperature. Given the assumed common scale
for all masses, we can calculate that they freeze out close to
their free-space masses and at the Braun-Munzinger et temperature,
and then drop all of these by 5 \% later. In this way, we are
doing our calculation as if the particles went back on shell
during the mixed phase. This applies more particularly to the
$\rho$-meson.

At $T_c$ each of the massless quark and antiquark coming together
to make up the $\rho$ will have the asymptotic energy $3.15T$, so
the $\rho$ is formed with energy $6.3T$. During the mixed phase
$T$ does not change, so it must have a mass of 770 MeV at the end
of the mixed phase where for purposes of calculation we assume it
to be on shell. In Table \ref{table1} we list the formation and
on-shell energies as function of temperature. We see from the
table that our assumption of $T_c\simeq T_{fo}$ works well, the
two being more or less equal at 177 MeV. Note that in NJL
calculations, $T_c$ is essentially unaffected by the bare quark
mass~\cite{bernard} and our asymptotic value of $6.3T$ should also
be negligibly affected.

\begin{table}
\caption{Formation and on-shell energies as function of
$T$}\label{table1}
\begin{center}
\begin{tabular}{ccc}
\hline
   $T$ (MeV) &$E_\rho^{th}$ in MeV  & 6.3$T$ \\
             & (with $m_\rho=770$ MeV) &  \\
\hline
 150 & 1070 & 945 \\
 175 & 1092 & 1102\\
 200 & 1145 & 1260\\
\hline
\end{tabular}
\end{center}
\end{table}

Our view of the phase transition is that the gluons turn into
massless vector mesons. How this can happen dynamically is
outlined in Appendix B which is most certainly not the complete
story. More theoretically, as proposed in \cite{BR:BERK}, one can
think of the gluon-meson conversion process as a Higgsing via the
color-flavor locking mechanism described by Berges and
Wetterich~\cite{berges-wett}. Although the color-flavor-locking
transition could very well be of first order as argued in
\cite{berges-wett} for the three-flavor case, in reality with
explicitly broken chiral symmetry with non-zero quark masses, we
see no obstruction to the notion that going off and on shell is
smooth as we are assuming here. We believe this is consistent with
the B\`eg-Shei theorem~\cite{beg-shei}~\footnote{There are related
ideas that come in different context which essentially say the
same thing; namely, Cheshire Cat Principle, gluon-meson duality
etc. This matter is discussed broadly in \cite{MR:taiwan}.}.  In
any case, there is a great similarity between massless vector
mesons and gluons, in that both are confined to the fireball. If
one imagines that color singlets of gluon pairs go back and forth
into pairs of vector mesons, the latter can only leave the system
when they are back on shell, although they can still scatter on
each other (with reduced vector coupling $g^*$ however) while
off-shell in the fireball. We will study the interactions of
off-shell vector mesons, as they turn into gluons and constituent
quarks below, using the lattice results for the quark number
susceptibility.

\begin{figure}[hbt]
\vskip 0cm
 \centerline{\epsfig{file=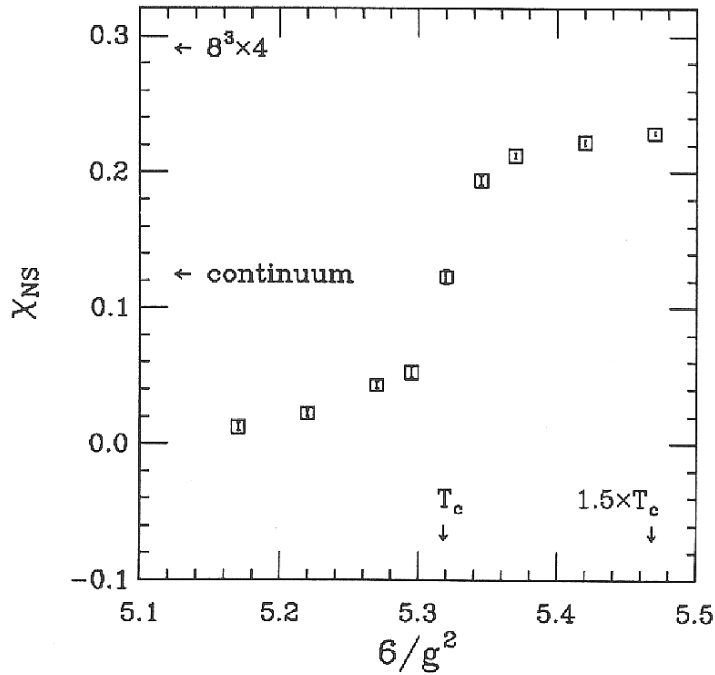,width=10cm,angle=0}}
\vskip 0cm \caption{\small The non-singlet (isovector) quark
number susceptibility measured by Gottlieb et al~\cite{gottlieb}
on lattice. The $g$ appearing here is the color gauge coupling of
QCD, denoted in this paper as $g_s$.}\label{chiNS}
\end{figure}

\begin{figure}[hbt]
\vskip -1.5cm
 \centerline{\epsfig{file=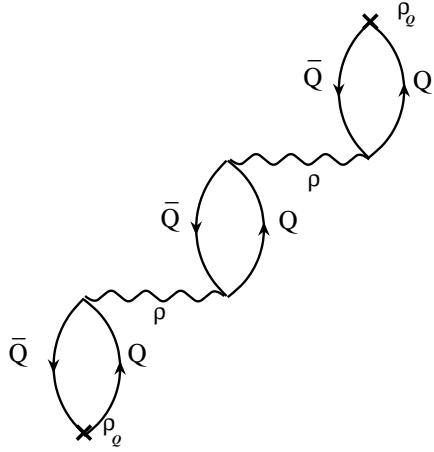,width=16cm,angle=0}}
\vskip -1.cm \caption{\small The quark number susceptibility below
the chiral transition temperature is described in RPA
approximation by summing quasi-quark-quasi-anti-quark bubbles by
exchange of $\rho$ mesons or, more properly, by four-Fermi (quark)
interactions in the vector-meson channel.}\label{bubble1}
\end{figure}

\begin{figure}[hbt]
\vskip -1.cm
 \centerline{\epsfig{file=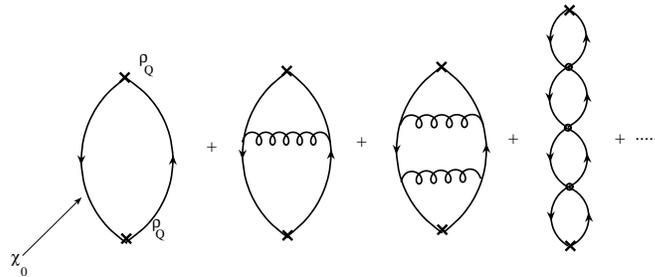,width=11cm,angle=0}}
\vskip -2.cm \caption{\small Perturbative calculation of the quark
number susceptibility above the chiral transition. The quark and
anti-quark coupled to the quark density $\rho_Q$ interact by
exchanging gluons depicted by the wavy line. The gluons are very
massive, so the wavy line should be shrunk to a point as indicated
by the graph with four bubbles. }\label{bubble2}
\end{figure}

In order to illustrate our principal point made above, namely,
that the gluons change smoothly into the vector mesons by the
Berges-Wetterich type relay, we consider the quark number
susceptibilities measured in lattice gauge
calculations~\cite{gottlieb}, the non-singlet one (in the $\rho$
channel) $\chi_{NS}$ of which is reproduced here in
Fig.\ref{chiNS}, and interpreted by Brown and Rho~\cite{BR:PR96}.
The description of these in RPA approximation is shown in
Fig.\ref{bubble1} in the hadronic sector below the chiral
restoration point~\footnote{It is interesting to note that this
RPA result in NJL -- which is a fermionic theory with four-fermi
interactions -- is reproduced by one-loop graphs of the bosonic
HLS/VM theory in \cite{HKRS}.} and Fig.\ref{bubble2} in the
quark/gluon sector above.
In the analysis of Brown and Rho~\cite{BR:PR96} the vector
coupling~\footnote{We reserve $g$ for the hadronic vector coupling
(e.g., HLS coupling). The color gluon coupling will be denoted,
whenever needed, as $g_s$, $\alpha_s$ etc.} of the $\rho$-meson to
nucleons at low temperature, $g^2/4\pi\sim 2.5$, gives way to that
of constituent quarks as the phase transition is approached and
then drops to $\alpha_s\approx 0.19$ (at $T=\frac 32 T_c$) after
the relay to the gluon side. In \cite{BR:PR96}, we conjectured
that this reflected the induced flavor symmetry {\em relayed} to
the fundamental color symmetry. This conjecture is supported by
the Berges-Wetterich scenario~\cite{wett,berges-wett}.

Let us study, using Fig.\ref{bubble1}, how the vector meson goes
on shell. We observe that the curve for the quark number
susceptibility calculated in lattice gauge calculation by Gottlieb
et al~\cite{gottlieb} (Fig.\ref{chiNS}) shows a smooth transition
through the critical temperature $T_c$. As in Brown and
Rho~\cite{BR:PR96} we can parameterize the susceptibility as
 \be
\chi (T)=\frac{\chi_0 (T)}{1+\frac 14 \frac{g^2}{m_V^2}\chi_0 (T)}
 \ee
where $\chi_0 (T)=2T^2$. For $T=0$, we have $g^2/4\pi=2.5$, so
that
 \be
g^2/m_V^2\approx 10\pi/m_V^2\approx 31/m_V^2.
 \ee
Now for $T=T_c$, $\chi\approx 0.5\chi_0=T_c^2$ from which we find
 \be
{g^\star}^2/{m_V^\star}^2\approx 2/T_c^2\approx 22/m_V^2
\label{attc}
 \ee
with $m_V\approx m_\rho$ and taking $T_c\approx 220$ MeV for the
quenched calculation of Gottlieb et al. It will turn out that the
precise value of $T_c$ is unimportant.

However, if we continue up to $\frac 32 T_c$, we see that
$\chi\approx .8\chi_0$ and we find
 \be
``g^2/m_V^2" \sim \frac{0.2}{T_c^2}.
 \ee
We indicated by the quotation mark an {\it effective} constant in
the quark-gluon sector that can be compared with the quantity in
the hadronic sector. We see that there is an order of magnitude
drop in this quantity in going from the hadronic into the gluonic
sector, roughly the same as in the $g^2/4\pi$-to-$\alpha_s$ drop.
Although the hidden gauge coupling $g$ and $m_V$ go to zero at
fixed points at chiral restoration, the ratio is finite and
continuous, showing that $g$ and $m_V$ scale in the same way as
predicted by the VM. More importantly, there is a smooth change in
$``g/m_V"$ from the QCD region where the $\chi$ should be
expressed in gluons as variables to that in which constituent
quarks should be employed as variables. The smoothness of this
quantity in the lattice gauge simulation of the susceptibility
also results from the relatively large bare quark masses used in
the simulation, giving a smooth crossover transition.

From the dimensional reduction calculation for the gluon side in
\cite{BR:PR96} we see that $\alpha_s\approx 0.19$, which grows to
$g^2/4\pi=2.5$ as the gluons are replaced by constituent quarks
and the $\rho$-meson goes back on shell. This may account for the
observation that transport calculations employing gluons with
gluonic couplings are unable to equilibrate sufficiently and
provide enough pressure for the HBT.

We remark that we would expect a similar behavior in $\chi$ as
function of density. The work of Song et al~\cite{songetal,song}
indicates that from zero density to $n_0$, the vector gauge
coupling $g$ remains roughly constant, but $m_V$ drops to
$m_V^\star (n_0)/m_V\simeq 0.8$. However, from then on to $n_c$,
$g^\star/m_V^\star$ remains more or less unchanged. The decrease
in $m_\rho^\star$ without change in $g$ was noted some time
ago~\cite{BRtensor} to increase the $\rho$-exchange contribution
to the tensor force in nuclei, thereby reducing the total tensor
force, since the $\rho$-exchange part has opposite sign to that
from the pion. This phenomenology supports the scenario that the
behavior of $g$ with density is similar to that with temperature.

\section{Conclusions}\label{conc}
\setcounter{equation}{0} 
\renewcommand{\theequation}{\mbox{\ref{conc}.\arabic{equation}}}
\itt In this paper, we have shown that the scale of Wilsonian
matching of constituent quarks to QCD is played by the
Nambu-Jona-Lasinio cutoff $\Lambda_4$. Introduction of the $\rho$
degrees of freedom changes the loop corrections in such a way that
the scale is changed to $\sqrt{2}\Lambda_4\simeq 4\pi f_\pi$, the
scale used by Harada and Yamawaki in their HLS theory.

The work of these authors gives a solid basis for BR scaling.
Indeed when the VM is implemented to NJL, we arrive at an overall
picture of how vector-meson masses -- both parametric and pole --
behave from low density to that of chiral restoration that is
consistent with, but rendered more precise than, the original BR
scaling. At very low density, chiral perturbation theory is
applicable with the $\rho$-meson treated as light and at densities
near chiral restoration, the vector manifestation provides a
reliable description. In between, our work presents the first
indication of how the (BR) scaling behavior interpolates. (This
scaling behavior has more recently been reproduced in a
description that is based on skyrmion structure~\cite{LPRV}). At
low densities, the parametric $F_\pi$ runs importantly while the
(HLS) gauge coupling $g$ stays more or less unchanged. So we
expect
 \be
m_\rho^\star/m_\rho\simeq f_\pi^\star/f_\pi\simeq
\left[\la\bar{q}q\ra^\star/\la\bar{q}q\ra\right]^{1/2}.
 \ee
This means that the phenomenological Lagrangian
approach~\cite{rapp} based on Lagrangians that have correct
symmetries is likely to be qualitatively correct up to $\sim n_0$.
However as $n$ exceeds $n_0$, the gauge coupling starts falling
locked to the VM fixed point and hence the phenomenological
approach that does not incorporate the flow to the VM fixed point
cannot correctly access the dynamics in this regime. It is the
HLS-type approach matched suitably to QCD that should take over.
As one approaches $n_c$, the vector meson (pole) mass should fall
linearly in $\la\bar{q}q\ra^\star$ with the width vanishing as
${g^\star}^2\sim (\la\bar{q}q\ra^\star)^2$.~\footnote{This feature
is encoded in the work of Harada, Kim and Rho~\cite{HKR} in the
following way. The $\rho$-meson pole mass is given by the
intrinsically running mass ${M^\star_\rho}^2=a^\star
{F^\star_\pi}^2 {g^\star}^2$ plus the dense-loop self-energy term
${g^\star}^2 R$ where $R$ is a slowly-varying function of density
or temperature. Up to $n\simeq n_0$, $g^\star$ does not scale, so
the intrinsic term dominates since the dense-loop corrections are
expected to be a small correction. Thus the $\rho$ mass scales
like $F^\star_\pi$ which scales as $(\la\bar{q}q\ra^\star)^{1/2}$.
Now for $n\gsim n_0$, the intrinsic mass $M_\rho$ scales as
$(\la\bar{q}q\ra^\star)^{3/2}$ since $g^\star$ scales as
$\la\bar{q}q\ra^\star$. Near the phase transition, therefore, the
intrinsic mass term drops off fast. Therefore the scaling of the
$\rho$ mass will be dictated by the dense-loop correction which
scales linearly in $\la\bar{q}q\ra^\star$.}

We find that for the vector manifestation the density for chiral
restoration is $n_c\lsim 5n_0$.

It is found that Rapp/Wambach ``widening" should be fused with
Brown/Rho scaling which corresponds to the intrinsic density
dependence inherent in the Wilsonian matching of the HLS effective
theory to QCD. It is likely that this fusing will lower $n_c$. In
fact, with the coupling between $\rho$ and $[N^\star
(1520)N^{-1}]^{1-}$ that Rapp and Wambach used, $n_c$ was
estimated to come down to $\sim 2.8n_0$~\cite{BLRRW}. In addition
given the difficulty in distinguishing between background and
subthreshold $\rho$ contribution, it is clearly too early to pin
down $n_c$. We can only say at this stage that $2.8 \lsim
(n_c/n_0)\lsim 5$ with the upper bound being perhaps more relevant
because of the explicit chiral symmetry breaking.

We have reinterpreted our long-standing argument that the
transition from the hadronic sector to the QCD sector at the
chiral transition point must be smooth in terms of the Higgsed
vector mesons shedding masses as density increases to de-Higgsed
(massless) gluons and similarly from gapped fermions (baryons) to
the gapless quarks etc. There is a compelling evidence from the
vector isospin susceptibility and axial-vector isospin
susceptibility that the massless vector mesons (or quasiquarks in
NJL) {\it must} figure as explicit degrees of freedom as do pions
near the chiral transition point~\cite{HKRS}. The presence of such
degrees of freedom for temperature near $T_c$ makes a completely
different prediction for the pion decay constants and pion
velocity from that of the pion-only scenario.

\subsection*{Acknowledgments}
\itt We are grateful for helpful discussions with Bengt Friman,
Masayasu Harada, Youngman Kim, Peter Kolb, Che Ming Ko, and Ismail
Zahed. This work was done while one of the authors (MR) was
spending three months in the Spring 2002 at the Theory Group, GSI
(Darmstadt, Germany) funded by the Humboldt Foundation. He thanks
GSI for the hospitality and the Humboldt Foundation for the
support.

\section*{Appendix A:  In the Beginning, the Color Glass}\label{app}
\setcounter{equation}{0} 
\renewcommand{\theequation}{\mbox{A.\arabic{equation}}}
\itt In Section \ref{RHIC} we presented a hadronization scenario
based on the Berges-Wetterich $relay$ mechanism, in which half of
the colored gluons get Higgsed~\footnote{Here we are concerned
with the $N_f=2$ case of \cite{berges-wett}.} as chiral symmetry
is broken in the expansion of the fireball formed in the heavy ion
collision.

In a series of papers, McLerran and Venugopalan have developed the
color glass model, summarized recently in \cite{krasnitz}. Here we
give a schematic model of how the colored glass is shattered,
skipping over the detailed picture of McLerran and Venugopalan but
focusing only on the hadronization scenario.

With a $\Gamma\sim 100$, RHIC produces matter originally at $n\sim
100 n_0$. However, only one unit of rapidity should be considered,
since different units don't communicate with each other.
Therefore, we have a co-moving $n\sim 20n_0$ per rapidity
interval. Now chiral restoration takes place in the time that it
takes the nucleon to communicate with the condensate which can be
described by the exchange of a scalar degree of freedom, say, a
$\sigma$, $\tau_{\chi R}\sim \hbar/(1\ {\rm GeV})=0.2$ fm/c --
called ``$\hbar/Q_s$" in the color-glass theory, if we include the
kinetic energy with unit mass of the $\sigma$. In this time,
because of the baryon density, the system will be chirally
restored, the high baryon density providing the important
collective effect which distinguishes the nucleus-nucleus
collision so markedly from a sum of nucleon-nucleon collision.

The central parts of the colliding sheets of colored glass are
shattered through the rapid chiral restoration in $\sim 0.2$ fm/c.
Here we will assume the same sort of curve for the susceptibility
$\chi_{NS}$ as function of density scale as temperature scale:
with chiral restoration, the couplings will change over from
hadronic to perturbative gluonic, thereby going to the right side
of Fig.2, with small $\alpha_s$ gluons. The saturation $\alpha_s$
of McLerran-Venugopalan is 0.5, somewhat larger than what we
arrived at in Section 6. The gluons are so copious in the
McLerran-Venugopalan theory that each momentum state has many of
them, and a classical effective field theory can be constructed.
Presumably at this stage one can talk about temperature because
the thermal energy is known to be $\sim (E^2+B^2)/8\pi$ assuming
equipartition.

The color glass is shattered into perturbative gluons at about 0.3
fm/c. Quarks are less important, being suppressed by order
$\alpha_s$ as compared with gluons. One can discuss the expansion
in terms of the entropy current $\tau\sigma$ where
$\tau=\sqrt{t^2-z^2}$ is the proper time, if the system is
equilibrated. Hydrodynamics~\cite{shuryak,heinz-kolb} shows that
in order to obtain sufficient flow, equilibration must set in very
soon following the color glass phase, at $\sim 1$ fm/c. The
subsequent behavior of the hot matter seems to be like that of
sticky molasses, rather than perturbative quark-gluon plasma.

The missing part of the RHIC scenario is how equilibration is
accomplished from the time of the breakup of the color glass down
to the hadronization. The large elliptic flow $v_2$ which develops
early tells us that the system does not behave like the predicted
weakly interacting QGP but more like sticky molasses. This
interpretation is supported by the low viscosity~\cite{teaney},
which implies short mean free path and by the behavior of the
``balance function"~\cite{pratt} in which the particles born in
the decay of a given particle do not get very far from each other
in rapidity in the time that the fireball evolves. Tantalizingly,
there are indications from the calculation of hadronic spectral
functions about the QCD phase transition in the maximum entropy
method (MEM) that there are strong collective excitations above
$T_c$~\cite{MEM}: Strong scalar, pseudoscalar, vector and
axial-vector hadronic modes are seen.

\section*{Appendix B: Higgsing of the Gluon; Dynamical Color Purifying
Filters}\label{higgsing}
\setcounter{equation}{0} 
\renewcommand{\theequation}{\mbox{B.\arabic{equation}}}
\itt While how the flavor gauge symmetry goes over to the color
gauge symmetry at the phase transition could be quite complex and
intricate depending on the number of flavors involved, non-zero
quark masses and a variety of collective phenomena (e.g., possible
meson condensation), our scenario is independent of the detailed
way in which gluons disappear with spontaneous break-down of
chiral symmetry and the energy goes into mesons in the central
rapidity (nearly baryon-free) region. The strongly interacting
mesons, such as the $\pi$, $\rho$ and $a_1$ as suggested in
\cite{BJBP} will equilibrate as they start to get their masses
back in going on-shell.

Although our scenario must happen on general grounds as stressed
before, it is nonetheless interesting that there are processes
that indicate that colored gluons do turn into colorless vector
mesons. For instance, colliding gluons can be turned into vector
meson pairs through coupling to $Q^2\bar{Q}^2$ states (where $Q$
stands as before for constituent quark) in conventional nuclear
physics~\cite{liuetal}. As we have argued in the main text that
the Rapp/Wambach, in conventional nuclear physics language, is
only part of the dropping mass story, the main part being the
parametric dependence of the vector meson on
density/temperature~\cite{BR91,HS:T,HKR}, the Li-Liu mechanism
that we refer to as ``dynamical Higgsing" may be only part of
Higgsing, the main part coming from confinement which forces color
to disappear with the chiral symmetry breaking (and might be
phenomenologically mocked up in terms of strongly increasing
attractions in color singlet states).

In the Li-Liu work, two off-shell gluons, one from the pion and
the other from the proton in a collision such as gives Drell-Yan
process, can be converted into two $\rho$-mesons, via a {\it
dynamical Higgsing} of gluons, into vector mesons. The two gluons
couple through $Q^2\bar{Q}^2$ (exotic) states considered a long
time ago in \cite{jaffe,liu-wong}. There are a number of these
states in the energy range $\gsim 1$ GeV and of particular
interest is an $I=2$ state at $\sim 1.6$ GeV, which is seen
experimentally, as well as $I=0$ states. The $I=2$ state can be
characterized as \be
 \Psi (Q^2\bar{Q}^2 (I=2))=a\sum_{\alpha=1}^8
 {V}^\alpha\cdot {V}^\alpha +bV.V\label{B1}
\ee where the $V^\alpha$ are colored vector mesons,
$\bar{q}\gamma_\mu\lambda^\alpha q$, and $V$ are the usual
colorless vector ($\rho$) mesons. The two gluons couple to $\Psi$
through the first term in the wave function with coupling constant
$\alpha_s$ under the assumption of colored vector dominance
adopted by \cite{liuetal} and decay into $\rho$-mesons through the
second term in the wave function.

This scenario is given credence by the behavior of the
$\gamma\gamma\rightarrow \rho^0\rho^0$ cross
section~\cite{liliu2}, to which the gluon-gluon-to-$\rho$-$\rho$
can be directly mapped. Experimentally a sharp peak in the cross
section is found at $E\sim 1.6$ GeV, with cross section of
$\sigma\sim 10^{-7}$ barns.  For gluons, $\alpha_s^2$ should
replace the electromagnetic fine structure constant squared,
$\alpha^2$, giving a factor of $\sim 5000$ for $\alpha_s\approx
0.5$. Thus, we estimate $\sigma \sim$  0.5 mb for two gluons to go
into two $\rho^0$'s.

There must be model states where the $V\cdot V$ piece in
(\ref{B1}) is replaced by $\pi\cdot\pi$ or $A\cdot A$ (where $A$
is the axial vector), so there may be substantial direct
production of the pions and axial vector mesons in the coupled
$\pi$, $\rho$, $a_1$ system, as well as formation from the vector
mesons by equilibration. Thus, our ``dynamical Higgsing" may have
a cross section as large as several mbs for gluons of the most
favorable energies

We expect the wave function (\ref{B1}) to survive chiral
transition without much change. The colored fields $V^\alpha$
presumably lie at a higher scale and the vector meson mean field
$V\sim g^\star/m^\star_V$ was shown above to increase $\sim 25\%$
in going from $n=0$ to $n=n_0$, and then remain constant through
chiral restoration. Thus, the $Q^2\bar{Q}^2$ states may, in fact,
do the dynamical Higgsing of converting gluons into vector mesons,
although they must be converted ultimately into systems of
strongly interacting mesons, with the latter then going on shell.

Incidentally, if the filters are not big enough to purify the
color, there may be some ``sludge" left in terms of glueballs. But
these will go out with expansion giving rise to decreasing
temperature because of small Boltzmann factors.

Now some caveats.

Although the group structure in the Higgsing and de-Higgsing can
be identified unambiguously, it is not feasible to  construct the
renormalization group flow from perturbative partons to hadrons,
although in our scenario the latter would begin massless in the
chiral limit. The problem is that in practice we are quite far
from the chiral limit, the constituent quarks having $\sim 100$
MeV masses in NJL as the dynamically generated mass goes to zero.
Thus, as seen in the susceptibility, Fig.\ref{chiNS}, the
transition is a smooth crossover one. We expect the Higgsing from
gluons to vector mesons to take place over this crossover.

\end{document}